# Consciousness as a physical process caused by the organization of energy in the brain

Robert Pepperell
Fovolab,
Cardiff Metropolitan University,
Cardiff, CF5 2YB, UK.
[rpepperell@cardiffmet.ac.uk]

**Abstract**

To explain consciousness as a physical process we must acknowledge the role of energy in the brain. Energetic activity is fundamental to all physical processes and causally drives biological behaviour. Recent neuroscientific evidence can be interpreted in a way that suggests consciousness is a product of the organization of energetic activity in the brain. The nature of energy itself, though, remains largely mysterious, and we do not fully understand how it contributes to brain function or consciousness. According to the principle outlined here, energy, along with forces and work, can be described as actualized differences of motion and tension. By observing physical systems, we can infer there is something it is like to undergo actualized difference from the intrinsic perspective of the system. Consciousness occurs because there is something it is like, intrinsically, to undergo a certain organization of actualized differences in the brain.

**Introduction**

> "If mental processes are indeed physical processes, then there is something it is like, intrinsically, to undergo certain physical processes. What it is for such a thing to be the case remains a mystery." (Nagel, 1974)

The philosopher Thomas Nagel summarised one of our greatest intellectual challenges: how to explain mental processes as physical processes. The aim of this paper is to outline a principle according to which consciousness could be explained as a physical process caused by the organization of energy in the brain.[1]

Energy is fundamentally important in all physical processes (Boltzman, 1886; Lotka, 1922; Schrödinger, 1944; Heisenberg, 1958). As the biophysicist Harold Morowitz put it: "the flow of energy through a system acts to organize that system" (Morowitz, 1979). Light, chemical reactions, electricity, mechanical work, heat, and life itself can all be described in terms of energetic activity (Chaisson, 2001; Morowitz and Smith, 2007; Smil, 2008) as can metabolic processes in the body and brain (Magistretti, 2008; Perez Velazquez, 2009). It is surprising, therefore, that energy receives relatively little attention in neuroscientific and psychological studies of consciousness. Leading scientific theories of consciousness do not reference it (Crick and Koch, 2003; Edelman et al., 2011;

---

[1] I take it that physical processes occur in time and space and are causally determined by the actions of energy, forces and work upon matter. I take consciousness to be the capacity for awareness of self and world, which is particularly highly developed in humans.

Dehaene, 2014; Oizumi et al., 2014), assign it only a marginal role (Hameroff and Penrose, 2014), or treat it as an information-theoretical quantity (Friston, 2013; Riehl et al., 2017). If it is discussed, it is either as a substrate underpinning higher level emergent dynamics (Deacon, 2013) or as powering neural information processing (Sterling and Laughlin, 2017).

This lack of attention is all the more surprising given that some of the pioneers of neurobiology, psychology, and physiology found a central place for energy in their theories, including Hermann von Helmholtz (in Cahan, 1995), Gustav Fechner (Fechner, 1905), Sigmund Freud (Gay, 1998), William James (James, 1907), and Charles Sherrington (Sherrington, 1940).[2] There are, however, signs that attention is turning again to energetic or thermodynamic-related theories of consciousness in various branches of science (Deacon, 2013; Collell et al., 2015; Annila, 2016; Tozzi et al., 2016; Street, 2016; Marchetti, 2018) and in philosophy of mind (Strawson, 2008/2017).

The present paper builds on this work by proposing that energy, and the related properties of force and work, can be described as *actualized differences* of motion and tension, and that — in Nagel's phrase — 'there is something it is like, intrinsically, to undergo' actualised differences. Recent neuroscientific evidence suggests that consciousness is a product of the way energetic activity is organized in the brain. Following this evidence, I propose that we experience consciousness because there is something it is like, intrinsically, to undergo a certain organization of actualized differences in the brain.

Several researchers have tackled the problem of consciousness by treating the brain in principle as a neural information processor (e.g. Tononi et al., 2016; Dehaene et al., 2017; Ruffini, 2017). I will argue that the governing principle of the brain at the neural level is not information processing but energy processing. The information-theoretic approach to measuring and modelling brain activity, however, can usefully complement the energetic approach outlined here.

**1. Consciousness and energy in the brain**

We do not fully understand the biological function of energy in the brain or how it relates to the presence of consciousness in the person.[3] Given that the human brain accounts for only 2% of the body's mass it demands a large portion of the body's total energy budget, some 20% (Laughlin, 2001; Magistretti & Allaman, 2013). Most of this energy is derived from the oxidization of glucose supplied to the cerebral tissue through the blood. Roy and Sherrington were the first to propose a direct correspondence between changes in cerebral blood flow and functional activity (Roy and Sherrington, 1890). Many features of human brain anatomy, such as the number of blood vessels per unit of space, the lengths of neural connections, the width of axons, and even the ratio of brain to stomach size are thought to be determined by the high metabolic demands associated with complex cognitive processing (Allen, 2009).

---

[2] For further discussion on the historical context see Pepperell (2018).
[3] Although for the sake of brevity I refer in this paper to consciousness occurring in the brain, consciousness is something that people undergo. Brains cannot sustain consciousness independently of the people in which they are housed (Pepperell, 1995/2003).

For many neuroscientists, the main function of energy in the brain is to fuel neural signalling and information processing (Magistretti, 2013); energy supply is seen as a constraint on the design and operation of the brain's computational architecture (Laughlin, 2001; Hall et al., 2012; Sterling and Laughlin, 2017). It has been calculated, for example, that the rate of energy supply available to the human brain places an upper 'speed limit' on neural processing of about 1 kHz (Attwell and Gibb, 2005). And Schölvinck et al. (2008) estimated that conscious perception of sensory stimuli increases energy consumption in primate brains by less that 6% compared to energy consumption in the absence of conscious perception.[4] They attribute this relatively small change to an energy efficient "design strategy" of the brain in which decreases in neural activity play a functional role in information processing as well as increases. Energy, on these accounts, plays no direct role in higher mental processes, like consciousness.

Robert Shulman and colleagues have argued there is a direct connection between energy in the brain and consciousness (Shulman et al., 2009; Shulman, 2013). By studying the progressive loss of behavioural response to external stimulus from wakefulness to deep anaesthesia, they found a corresponding reduction and localisation of cerebral metabolism (a marker of energy consumption). Therefore, they argue, high global metabolism is necessary for consciousness. However, they are also clear that high global metabolic rates are not sufficient as patients with locked-in-syndrome and those who suffer from some forms of epileptic seizure can register high levels of global brain metabolism without exhibiting the observable behaviour that we expect from a conscious person (Shulman, 2013; Bazzigaluppi et al., 2017). Shulman's thesis has been challenged on several grounds (Seth, 2014). For example, it has been pointed out that behavioural responsiveness may be inadequate as a measure of sentience given that vestiges of consciousness have been detected in people diagnosed as being in a vegetative state with a low cerebral metabolism (Owen et al., 2006). Moreover, some patients who recover from a vegetative state to regain consciousness do so despite having substantially reduced cerebral metabolism compared with normal controls (Laureys et al., 1999; Chatelle et al., 2011).

In recent years there has been a growing interest in intrinsic brain activity (Clarke and Sokoloff, 1999; Raichle, 2011). This background or spontaneous activity occurs in the resting awake state in the absence of external stimulation or directed attention, and its energy demands can greatly exceed those of localised activation due to task performance or attention. The discovery of this so-called 'dark energy' in the brain (Raichle, 2010) was greeted with some surprise in the neuroscience community and remains controversial (Morcom & Fletcher, 2007). Work on intrinsic activity led to the identification of a 'default mode network' in the brain, an extended set of interconnected regions that uses high levels of energy when a person is in a non-attentive state. Energy use drops significantly in this network when a more cognitively demanding task, such as paying attention to a stimulus, is performed (Shulman et al., 1997; Raichle et al., 2001). Vanhaudenhuyse et al. (2009) reported that connectivity within the default mode network in patients with severe brain-damage deteriorates in proportion to the degree of conscious impairment, suggesting it plays an important role in sustaining consciousness.

Meanwhile, it is somewhat surprising to find that energy use during non-rapid eye movement sleep remains at ~85% of that in the waking state, while during rapid eye movement sleep it can be as high as in the waking state (Dinuzzo et al., 2017). At the same time, consciousness can be minimally sustained with energy use at only 42% of the level that occurs in healthy conscious individuals, suggesting that much cerebral metabolic activity in normal waking states does not directly contribute to consciousness (Stender et al., 2016). Many anaesthetic agents are thought to obliterate consciousness because they reduce the global rate of cerebral metabolism (Hudetz, 2012). Administering ketamine, on the other hand, increases brain metabolism yet can still lead to loss of responsiveness (Pai and Heining, 2007). Overall, it seems we find no clear correlation between the total amount of energy used by the brain, or the location where the energy is used, and the level of consciousness detectable in the person.

---

[4] Strictly speaking energy is not consumed but converted from one form to another.

## 2. Consciousness and the organization of energetic processing in the brain

An alternative, or perhaps complementary, way to think about this issue is in terms of how the energetic activity in the brain is organized rather than its global level or localisation. Indeed, this has implicitly been the focus of recent research that aims to provide quantitative measures of consciousness levels. In one study, researchers used transcranial magnetic stimulation (TMS) to send a magnetic pulse through the brains of healthy controls and patients with various states of impaired consciousness (Casali et al., 2013). By measuring how the pulse perturbed the cortex the researchers were able to determine the relative complexity and extent of the pathways through which the pulse propagated and correlate these to levels of consciousness. The researchers calculated a perturbation-complexity index (PCI) that quantified the levels of consciousness present in each person they studied. This method was further validated as a reliable objective measure of levels of consciousness by Casarotto et al. (2016).

The PCI was calculated using data from electroencephalographic (EEG) measurements of the cerebral perturbation following the TMS. Images from the EEG were filtered into binary data that was then analysed using a Lempel–Ziv algorithm, a commonly used information-theoretical technique in which complexity is measured as a function of data string compressibility, with more complex data strings being less compressible (Ziv and Lempel, 1977; Aboy et al., 2006). Other researchers have developed similar information-theoretical methods for quantifying the complexity of brain activity and levels of consciousness. King et al. (2013) analysed data from 181 EEG recordings of patients who were diagnosed with varying states of impaired consciousness and applied a measure of weighted symbolic mutual information (wSMI) that sharply distinguished between patients in vegetative state, minimally conscious state, and conscious state.

Although information theoretic tools were being used to analyse and interpret the data in these studies we should note that what was actually being detected by the experimental procedures was not information per se but the organization of energetic activity or processing in the brain. Energetic processing — the processes by which the brain regulates the flow of energy through its structures — is routinely detected at varying degrees of spatial and temporal resolution, either directly or indirectly, by neuroimaging techniques such as positron emission tomography (PET), functional magnetic resonance image (fMRI) and EEG (Shulman, 2013; Bailey et al., 2005; Niedermeyer and Lopes da Silva, 1987). Referring again to the study by Casali et al. (2013), the perturbations from which the PCI was calculated were generated by a pulse of magnetic energy from the TMS and were imaged with EEG that measures electrical voltage differences, that is, fluctuations in energetic potentials between clusters of neurons in the cortex (Niedermeyer and Lopes da Silva, 1987; Hu et al., 2009; Koponen et al., 2015). The PCI and wSMI can therefore be interpreted as measures of the complexity or organization of energetic processing in the brain during the experimental procedures.

Subsequent research has directly investigated the connection between brain metabolism (how the brain regulates energy conversion), brain organization, and levels of consciousness by combining EEG measures with PET, a more specific measure of cerebral metabolism. Chennu et al. (2017) collected data from 104 patients in varying states of conscious impairment using both techniques. By analysing this data, they determined a metric that discriminated levels of consciousness to a high degree of accuracy. This study built on previous work by Demertzi et al. (2015) that used fMRI to correlate a measure of intrinsic functional connectivity in the brain with levels of consciousness. The PCI method has been further validated by a study combining EEG and $^{18}$F-fluorodeoxyglucose (FDG)-PET (Bodart et al., 2017), so reinforcing the link between levels of consciousness and the organization of metabolic activity in the brain.

Current brain imaging methods do not strictly speaking detect information processing.[5] They do, however, detect changes associated with increases in energy consumption (via fMRI and PET) and fluctuations in electrical potential energy (via EEG), both of which reliably correlate with changes in mental function and behaviour. On the basis of what we can observe, the brain operates according to the principle of energetic processing. The evidence discussed above suggests levels of consciousness are determined by the organization of energy processing in the brain rather than on its global level or localization; wakeful conscious states are associated with more complex organization. To understand why this might be we need to consider the concept of energy in more depth.

## 3. Energy

The concept of energy that we are familiar with today emerged only slowly from its beginnings in the late eighteenth century. It developed through the study of thermodynamics in the nineteenth century, and then found its place at the centre of theories of relativity, quantum mechanics, and cosmology in the twentieth (Coopersmith, 2010). In colloquial usage energy refers to ideas of vigour, vitality, power, activity, and zest. In scientific usage, however, energy is defined as the ability of a system to do work.[6] Work is defined as the transfer of energy involved in moving an object over a distance by an external force, at least part of which is applied in the direction of the displacement (Duncan, 2002). Scientists and engineers often refer to energy as an abstract property: "Energy is a mathematical abstraction that has no existence apart from its functional relationship to other variables" (Abbott and Van Ness, 1972. See also Rose, 1986). It is a property that can be converted from one form to another, and in an isolated system the total quantity is conserved (Smil, 2008).

Despite the enormous amount of interest in the physics of energy and its central importance in so many branches of science, its nature remains in many ways mysterious (Feynman, 1963; Smil, 2008; Coopersmith, 2010) and it has been the subject of relatively little philosophical interrogation (Coelho, 2009). Treating energy as an abstract accounting quantity is perfectly satisfactory for many scientific purposes, where there is little reason to question its nature. But if energetic activity plays a significant role in consciousness, as the evidence cited above suggests it might, then its nature deserves closer scrutiny.

---

[5] The authors of Wollstadt (2017), for example, studied the breakdown of local information processing under anaesthesia using information theoretic methods. They point out that the EEG procedure they used did not directly record information processing in the brain but local field potentials, that is, fluctuations in quantities of potential energy.

[6] There seems to be an ambiguity in some textbooks about whether energy is an enabling property possessed by a system or body, e.g. Duncan (2002), or a measure of such a property, e.g. Rennie (2015). I will take energy to be a property possessed by systems or bodies, quantities of which can be measured.

The concept of energy in the European intellectual tradition can be traced back to Aristotle who used but never precisely defined the term *energeia* (ενέργεια) to convey the ideas of action, activity, actuality, being at work, and acting purposefully (Sachs in Aristotle, 2002). Scholars have long debated the best way to translate *energeia* from ancient Greek. The word 'energy' itself has been used, as have 'activity' and 'actuality', but 'being-at-work' is currently favoured, partly due to *energeia's* roots in *ergon*, the ancient Greek for work (Aristotle, 1818; Ellrod, 1982; Sachs in Aristotle, 2002). Modern scholars have tended to quarantine the ancient concept of *energeia* from contemporary ideas about energy. But Aristotle's term may still have value when thinking about energy's nature. This is especially so when we take into account the ideas of Aristotle's intellectual forebear Heraclitus, whose cosmological view was informed by three main principles: (i) that activity in nature is driven by 'fire' — which has been interpreted as synonymous with energy (Heisenberg, 1958), (ii) is subject to continual flux or motion, and (iii) is structured by antagonism or tension and (Kahn, 1989; Sachs in Aristotle, 2002).

We now understand there to be two main forms of energy: kinetic and potential. Kinetic energy is possessed by objects due to their motion, while potential energy is possessed by objects due to their relative position or configuration. All other forms of energy, such as thermal, electromagnetic, solar, chemical, gravitational, atomic and so on are in themselves forms of either kinetic or potential energy (Duncan, 2002; Smil, 2008). Much can be said about kinetic and potential energy, including the fact that they are causally efficacious, that is, they cause real change and activity in the material world.[7] But I want to draw attention here to the fact that they are both expressions of *difference*. Kinetic energy is difference as motion or change; potential energy is difference as tension or antagonism. Neither kinetic nor potential energy inhere absolutely in objects but are relational properties; motion or change occurs relative to a frame of reference, and tension or antagonism ocuurs between one object, or force, and another. The concept of *difference* then is of utmost importance when considering the nature of energy and the related properties of force and work.[8]

## 4. Actualized difference

If energy is the ability to do work then the displacement of a body undergoing work is due to force, defined as the "agency that tends to change the momentum of a massive body" (Rennie, 2015) or less formally as a "push or a pull". Forces act and react antagonistically in equally opposing pairs and are therefore, like energy, expressions of difference. The discipline of physics finds it convenient to treat energy, forces and work as distinct quantities to be balanced in abstract mathematical equations. But in nature they are integral and actualized, acting collectively in time and space with causal efficacy.

---

[7] "Energy may be called the fundamental cause for all change in the world" (Heisenberg, 1958). The neurobiologist Gerald Edelman neatly defined causal efficacy as "The action in the physical world of forces or energies that lead to effects or physical outcomes" (Edelman, 2004).

[8] Neuroanthropologist Terence Deacon defines energy as a "relationship of difference" (Deacon, 2013). Note that energy is difference, but not all differences are energy; red is a colour, not all colours are red.

By observing nature, we can infer there is 'something it is like' to be a physical system undergoing antagonistic forceful interactions, and what it is like will vary as the interactions vary.[9] There is something it is like, for example, to be a piece of rope undergoing great tension that is different from what it is like to be the same rope when relaxed, or for a rock to crash to earth having been in freefall. Some effects of these interactions may be observed from an extrinsic perspective; we may hear a creak or a crunch. But the something it is like to undergo the interactions themselves is an intrinsic property of the observed system to which the extrinsic observer has no access. It is for this reason that its presence and nature can only be inferred.[10]

This is not to claim that forces acting at the subatomic scale between particles, or at the macro scale in rope or rock, undergo anything like the experience we undergo as conscious humans.[11] Something it is like-ness is not in itself consciousness. Rather, it is to recognise that:

(i)     energy, forces, and work are *actualized*,
(ii)    they are expressions of *difference,* and
(iii)   there is *something it is like*, intrinsically, to undergo actualized difference.

I use the term *actualized difference* to refer to the active, antagonistic nature of energy, forces and work in a way that encompasses Heraclitean cosmology, Aristotlean *energeia,* and contemporary scientific descriptions of energy. Mathematical equations can represent actualized differences with abstract differences, in the form of symbols and numbers, but not whatever it is that puts the "fire in the equation" (Hawking, 1988).[12] For that we must refer back to nature itself — to the territory rather than the map (Korzybski, 1933).

**5. Energy and information**

For many contemporary scientists, information is a fundamental property of nature. For some it is *the* most fundamental property of nature (Davies, 2010). Neuroscientists often claim that the brain operates according to the principle of information processing. We read that "the brain is fundamentally an organ that manipulates information" (Sterling and Laughlin, 2017) and that brains are "information processing machines" (Ruffini, 2017). Individual neurons are treated as information processing units, while neural firing patterns are converted into sequences of binary digits (1s and 0s) that encode information (Koch, 2004). Recent prominent theories claim consciousness is identical with (Tononi et al., 2016) or results from (Dehaene et al., 2017) certain kinds of information structures or information processes in brains.

---

[9] Nagel clarified the term 'something it is like' as meaning not what something resembles but 'how it is' for the system (Nagel, 1974).

[10] Note that this claim is not as far-fetched as it might at first seem: If (i) consciousness in people is a physical process — due to energy, forces and work — and (ii) we infer the presence of consciousness in other people on the basis of observing them extrinsically — as we habitually do — and (iii) there is something it is like to be a conscious person — as we assume there is — then (iv) we routinely infer the presence of an intrinsic something it is like-ness in a physical process on the basis of observing it from an extrinsic perspective. However, as discussed below, human consciousness is a particular kind of something it is like-ness that occurs only when certain conditions are met.

[11] In discussions of the nature and behaviour of forces at the microscopic level we often find references to the way they 'feel' (Feynman, 1963), or the way they 'experience' each other in fields (Rennie, 2015). It would be interesting to investigate what motivates the use of such terms in this context.

[12] The difference between 1 and 0, for example, is an abstract difference conceived within a conscious mind.

Information is variously and sometimes imprecisely defined in science (Capurro and Hjørland, 2005), its meaning is still strongly contested (Lombardi et al., 2016; Roederer, 2016), and many people regard it as being to some extent subjective, relativistic, or observer-dependent (von Foerster, 2003; Deacon, 2010; Werner, 2011; Logan, 2012; Searle, 2013; de-Wit et al., 2016). The term is often used in science colloquially (meaning 'what is conveyed by an arrangement of things') or "intuitively" (Erra et al., 2016). And where one might expect to find a clear definition, such as in a dictionary of physics, biology or chemistry, none appears (Rennie, 2015; Hine, 2015; Rennie, 2016).

The most widely cited technical definition of information is that given by Claude Shannon (1948) as part of his mathematical theory of communication. For Shannon, information does not refer to meaning or semantics, as it does colloquially. The information is the amount of uncertainty in a message (a sequence of data) measured through probabilistic analysis of its elements. Information theory has developed into an exceptionally powerful mathematical tool that can be used, among many other things, to measure the complexity of physical systems. But a quantity of Shannon information is a measure of what can be *known* about a system as distinct from the system itself. The information lies with the measurer rather than the measured.[13]

The other commonly cited definition of information is Gregory Bateson's "a difference that makes a difference" (Bateson, 1979). Like his fellow cybernetic theorist Norbert Wiener (1948), Bateson sharply distinguished information from energy. Difference is not a property of what he calls the "ordinary material universe" governed by energetic activity. It is not subject to the effects of impacts and forces but is an abstract, relational property of the mind that exists outside the realm of physical causation: "Difference, being of the nature of relationship, is not located in time or space". Information defined according to Bateson as a "nonsubstantial" abstract difference cannot be used to explain consciousness as a physical process.[14]

The integrated information theory of consciousness (IIT) proposed by Tononi and colleagues provides an alternative, non-Shannonian, definition of information as "a form in cause-effect space" (Tononi et al., 2016). Cause-effect space, according to their theory, contains a "conceptual structure"— a constellation of related concepts — that is specified by the "physical substrate of consciousness" (PSC), this being the precise complexes of neural activation involved in any experience. Each conscious experience is identical with this "form", denoted $\Phi^{max}$ when maximally integrated. But while IIT is presented as a theory of integrated information, it could equally serve as a theory of how energetic processing is organized since the PSC consists in the causally interrelated patterns of neural firing that are identical with the conscious experience.

---

[13] Arieh Ben-Naim sets out in some detail how Shannon information is a probabilistic measure rather than a physical property (Ben-Naim, 2015). Note that the act of measurement presupposes a conscious mind capable of carrying out the measurement procedure and interpreting the result.

[14] Had he a fuller understanding of the nature of energy Bateson might not have been so dismissive about its role in mental processes. In *Mind and Nature* (Bateson, 1979) he referred only to kinetic energy (which he defined as "$MV^2$"), thus ignoring potential energy, and was by his own admission "not up to date in modern physics". In fact, slightly modifying Bateson's much-cited phrase to *an actualized difference that makes a difference* yields a description of the essence of energetic action, that is, the way energy, forces and work act antagonistically to effect change and cause further actions.

Treating brains as neural information processors does not help us to understand consciousness as a physical process because information, according to the commonly accepted definitions, is not a physical property of brains at the neural level; *there is no information in a neuron.*[15] It is useful, however, to apply information-theoretical methods to study the organization of physical systems, such as brains. Norbert Wiener (1948) stated: "…the amount of information in a system is a measure of its degree of organization…" As exemplified in several studies and theories cited here, we can measure and model the way the organization of energetic processes in the brain contributes to the presence of consciousness in a person.[16] But the abstract difference between 0 and 1 is not equivalent to the actualized difference between a neuron at rest and firing.

## 6. The brain as a 'difference engine'

The challenge of explaining consciousness as a physical process is made more tractable, I suggest, by recognising that brains operate on the principle of energetic processing. Neurons, in concert with other material structures such as astrocytes and mitochondria, convert, distribute, and dissipate electro-chemical energy through highly organized pathways in order to drive behaviours critical to the organism's survival. This makes sense when we consider the fact that organisms inhabit a physical world that is structured through the actions of energy, forces and work. To survive and prosper in this world they must continually work to acquire new supplies of high-grade or free energy to maintain an internal state far from thermodynamic equilibrium (Boltzmann, 1886; Schrödinger, 1944; Schneider & Sagan, 2005). Besides internal regulation, nervous systems enable organisms to perform two major tasks: *discriminating* between variations in environmental conditions, such as temperature, acidity, salinity, nutrient levels, or presence of predators, and *moving* towards environmental conditions that are beneficial to survival and away from those that are harmful.

The mechanisms that enable performance of these tasks can be seen at work in organisms with relatively simple nervous systems, such as the C. elegans worm (Sterling and Laughlin, 2017). Chemical gradients in the environment activate chemosensory neurons on the worm's surface that connect via interneurons to motor neurons that control the action of dorsal and ventral muscles, which, in turn, control the worm's movement (de Bono and Maricq, 2005). In this way, differences of chemical potential energy in the environment are converted into differences of electro-chemical energy in the sensing apparatus of the organism and then into differences of chemical energy in the muscles, which, by antagonistic action, are converted into the kinetic energy of the organism's movement. The organism makes discriminations in the environment relevant to its interests so that it can take appropriate actions in response.

---

[15] Brains — as parts of people — process information in the colloquial sense, just as they process abstract ideas, equations, numbers, thoughts, emotions, or memories. But they do so as a *consequence* of the underlying energetic processing (conversion, distribution, dissipation) going on in neural tissue. Computers also 'process' information in the colloquial sense. Mechanically and electronically speaking, however, they actually manipulate energy states (voltages, light, etc.) the results of which we, as conscious people, interpret informationally. It is worth noting that all mechanical information processing necessarily entails the dissipation of a certain amount of energy (Landauer, 1961). Recent experiments have confirmed this principle and demonstrated the intimate link between energy and what many refer to as information (Bérut et al., 2012).

[16] Logan (2012), in work undertaken with Stuart Kauffman and others, defines 'biotic information' as the organization of the exchange of energy and matter between organism and environment — a further example of information theory being used to quantify the biological organization of energy flows.

We can see the same basic principle at work in biology of far greater complexity. The human visual system, for example, is highly demanding on the brain's energy budget (Wong-Riley, 2010). But the evolutionary benefit of human vision is the capacity it confers to guide finely controlled bodily actions in light of environmental conditions. This is achieved through an intricate sequence of energy conversions, beginning with the arrival at the retina of electro-magnetic energy from the environment and cascading through numerous energetic exchanges in the neural pathways of the visual system that progressively differentiate features of the environment (Hubel and Livingstone, 1987). This frequently results in the conversion of electro-chemical energy in the motor system and muscles to the kinetic energy of bodily movement (Goodale, 2014). The fact that our complex biology supports so rich a repertoire of sensory discriminations and motor responses may be only a difference of degree rather than of kind with the humble worm.

We might think of sensory cells responding to stimulation from environmental energy by becoming excited or by increasing local neural activation. But vertebrate photoreceptors are, contrary to what one might expect, hyperpolarised by photon absorption. This means they 'turn off' when exposed to light and 'turn on' in the dark, even though they use more energy in the dark (Wong-Riley, 2010).[17] Meanwhile, some of the neurobiological evidence cited in Section 1 cautions us against assuming that sensory stimulation always results in increased neural activation. Decreases in activation in the brain can occur in response to cognitively demanding tasks, yet can go unnoticed in imaging studies with methodologies designed to detect task-evoked increases in metabolic rate above baseline (Raichle et al., 2001; Schölvinck et al., 2008). And of course not all neural activation is excitatory; neural inhibition is vitally important in brain function, as elsewhere in the nervous system, and this also entails an energetic cost (Buzsáki et al., 2007). There is evidence that an optimal balance between neural excitation and inhibition (E-I balance) in the cerebral cortex is critical for the brain to function well (Zhou & Yu, 2018).

In light of these mechanisms, the energy-hungry brain might be understood as a kind of 'difference engine' that works by actuating complex patterns of motion (action potential propagation) and tension (antagonistic pushes and pulls between forces) at various spatiotemporal scales. Firing rates and electrical potentials vary within neurons, between neurons, between networks of neurons, and between brain regions, so maximising the differential states the brain undergoes. A decrease in activation, or a reduction in firing rate, can create a differential state just as much as an increase. And, as is indicated by the work of Schölvinck et al. (2008), deactivation may be an energy efficient way for the brain to increase its repertoire of differential states. Maintaining a global E-I balance across spatiotemporal scales, meanwhile, is thought to promote 'efficient coding' in sensory and cognitive processing (Zhou & Yu, 2018). All this lends support to the idea, proposed above, that one of the roles of energetic activity in the brain is to efficiently actuate *differences* of motion and tension that advance the interests of the brain-bearing organism. It is the *actualized difference* that makes the difference.

---

[17] It turns out this arrangement is energy efficient for the visual system overall (Wong-Riley, 2010).

## 7. Energetic organization as the cause of consciousness

In theory, we could account for all the highly complex processes occurring in the brain in terms of energy, forces and work, that is, as physical, chemical and biological processes. But the seemingly unassailable problem of how any of these processes might cause consciousness remains. The principle outlined here — that there is something it is like, intrinsically to undergo differences due to the antagonistic action of energy, forces and work — may offer a toehold in the slippery face of the problem. There is something it is like, intrinsically, to be a tense muscle that is different from being a relaxed muscle. There is something it is like, intrinsically, to be networks of neurons in fantastically complex states of actualized differentiation from other networks, with action potentials propagating through vast arrays of fibres. But all this something is it like-ness is not in itself consciousness. Muscles are not conscious, and networks of neurons are active in the brain when we are in dreamless sleep or under anaesthesia. What is it about the organization of energetic processes in the brain, as discussed in Section 2, that determines the level of consciousness we experience?

We gain some insight into the association between consciousness and the organization of energetic processing in the brain from studies of anaesthesia. The reason why anaesthetic agents obliterate consciousness is not understood (Mashour, 2004). Recent work has focused on the ways in which they interfere with the brain's capacity to generate patterns of localised differentiation (often termed 'information') and to bind together or integrate those patterns across widely distributed brain networks (Hudetz, 2012; Hudetz & Mashour, 2016). Evidence from studies on the neurological effects of anaesthetics suggests that consciousness is lost as distant regions of the brain become functionally isolated and global integration breaks down (Lewis et al., 2012). The idea that consciousness depends on maximising differentiation and integration in the brain lies at the heart of IIT (Tononi, 2012; Oizumi et al., 2014).

A potential mechanism supporting global integration of local differentiation is recurrent or reentrant processing, in which widely distributed areas of the brain engage in complex loops of cortical feedback via massively parallel connections (Edelman et al. 2011; Edelman & Gally, 2013). A number of studies of the effects of anaesthetics have shown that they disrupt feedback connectivity, and hence integration, particularly in the frontoparietal area of the brain (Lee et al., 2009; Hudetz & Mashour, 2016). Studies of brain organization during deep sleep have also reported an increase in modularity consistent with the loss of integration among regions of the brain found in the awake state (Tagliazucchi et al., 2013). This suggests that the presence of consciousness in a wakeful person depends on a certain level of functional integration supported by cortical feedback loops (Edelman, 2004; Alkire et al., 2008) but it is not known how or why.

A major contribution of cybernetic theory was to recognise the importance of feedback mechanisms for controlling behaviour in mechanical and living systems (Wiener, 1948; Bateson, 1972). Feedback systems are self-referential; behaviour of one part of the system casually affects another, which in turn affects the first. Such systems are apt to generate patterns of behaviour that are an irreducible property of the system as a whole (Hofstadter, 2007; Deacon, 2013). One example is video feedback, which occurs when a video camera is pointed at a screen showing the output from the camera (Crutchfield, 1984). When correctly arranged the screen will at first display a tunnel-like image that will then spon-taneously 'blossom' into an intricate, semi-stable pattern of remarkable diversity and fascinating beauty (see Figure 1).[18] Since this is an energetically actuated process we can infer, following the arguments already given, that there is something it is like to be the video feedback system in full bloom, from its intrinsic perspective.

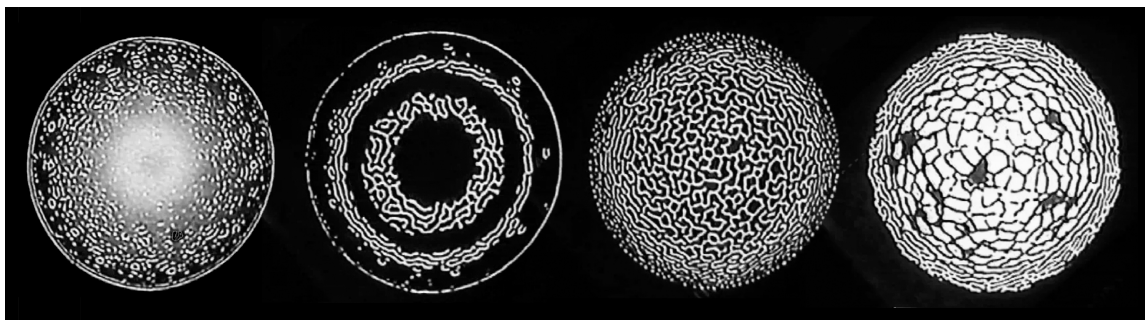

**Figure 1**. Stills from a video feedback sequence generated by the author. These patterns are created by pointing a video camera at a screen showing the camera's output. What begins as a tunnel-like images soon 'blossoms' into an ever-changing pattern of great diversity and fascinating beauty. © Robert Pepperell, 2018.

Gerald Edelman proposed that "phenomenal experience itself is entailed by appropriate reentrant intracortical activity" (Edelman & Gally, 2013). The human brain undergoes recursive or reentrant behaviour of an unimaginably higher order of complexity than in the video system.[19] But the underlying operating principle may be analogous. Video feedback arises because the system is organized as a self-observing loop. If we assume that reentrant activity in the brain is also a kind of self-observing loop in which processes in one part the brain both affect and are affected by processes in other parts, then we can envisage a kind of pattern blooming in the brain analogous to that we see in video feedback. This pattern would be actuated by sufficiently organized electro-chemical activity, among neurons and neurotransmitters, channelled through reentrant neural circuits.

The something it is like-ness a brain organized in this way would be undergoing is of a different order to that of a brain with diminished integration in dreamless sleep or under anaesthesia. No other physical system, as far as we know, has the same capacity for complex (differentiated and integrated) recursive processing as the human brain, and that dynamic organization reaches its apogee when we are wakefully conscious, as suggested by the evidence cited in Section 2. When the energetic processes in our brains are operating at a certain level of dynamic recursive organization — the "appropriate reentrant intracortical activity" — then we undergo something it is like, intrinsically, to undergo something it is like, intrinsically, to undergo something it is like … recursively. In other words, *there is something it is like, intrinsically, to be something it is like, recursively, to undergo the particular organization of actualized differences found in the conscious brain*. For this we have the most direct and irrefutable evidence possible — what it's like to undergo our own conscious experience.[20]

---

[18] Examples can be found on YouTube.
[19] One way to quantify this relative complexity would be to follow the proposal of Chaisson (2001) and compare the energy rate density (a measure he calls $\Phi_m$) between the two systems. Note also that Edelman and Gally are careful to distinguish cybernetic feedback in machine control systems from re-entrant processing in the brain, the latter being far more complex (Edelman & Gally, 2013).
[20] 'I think therefore I am undergoing a certain recursive organization of actualized differences.' Models of consciousness based on feedback loops in the brain have been discussed before, including by Douglas

Is it reasonable then to propose that consciousness is *caused* by the way energetic activity is dynamically and recursively organized in the brain? It is no less reasonable than attributing the causes of other biological phenomena, such as the behaviour of the nematode worm, to the way energetic activity is organized. If consciousness is a physical (biological and chemical) process, and if physical processes are caused by energetic activity (alongside forces and work), then consciousness, in principle, could be caused by energetic activity and the way it is organized.

**8. Naturalising consciousness**

In 1937–8 Charles Sherrington gave a series of lectures on the relationship between energy and mind, collected in the volume *Man on his Nature* (Sherrington, 1940). Drawing on the physics of his day, Sherrington understood the natural world to be composed of forms of energy. But he could not conceive how the mind could be forged from energy: "The energy-concept of Science collects all so-called 'forms' of energy into a flock and looks in vain for the mind among them." The mystery was deepened for him by the knowledge, then emerging through studies of electrical and metabolic activity in the brain, of how intimately energy and the mind must be linked. He was compelled to wonder "Is the mind in any strict sense energy?" but reluctantly concluded that "…thoughts, feelings, and so on are not amenable to the energy (matter) concept." They lie beyond the purview of natural science, despite the "embarrassment" this causes for biology.

If we are to naturalise consciousness, then we must reconcile energy and the mind. I have outlined a principle that may help to explain consciousness as a physical process. It entails re-examining the modern scientific concept of energy in the light of Aristotle's *energeia* and its Heraclitean roots. Accordingly, we arrive at a view of physical processes in nature as actualized differences of motion and tension. Sherrington understood that "Energy acts, i.e. is motion." But he went on "…of a mind a difficulty is to know whether it is motion." Treating the brain as a difference engine the work of which is to actualize and organize differences of motion and tension that serve the interests of the organism is, I submit, a natural approach to understanding consciousness as a physical process.

**Conclusion**

If consciousness is a natural physical process then it should be explicable in terms of energy, forces and work. Energy is a physical property of nature that is causally efficacious and, like forces and work, can be conceived as actualized differences of motion and tension.

Evidence from neurobiology indicates that the brain operates on the principle of energetic processing and that a certain organization of energy in the brain, measured with information theoretic techniques, can be reliably predict the presence and level of consciousness. Since energy is causally efficacious in physical systems, it is reasonable to claim that consciousness is in principle caused by energetic activity and how it is dynamically organized in the brain.

---

Hofstadter in his book *I am a Strange Loop* (Hofstadter, 2007). I have also previously proposed a feedback model of consciousness partly inspired by Edelman's theory of reentrant processing as part of an attempt to design an artificially conscious work of art (Pepperell, 2003).

Information in the biological context is best understood as a measure of the way energetic activity is organized, that is, its complexity or degree of differentiation and integration. Information theoretic techniques provide powerful tools for measuring, modelling, and mapping the organization of energetic processes, but we should not confuse the map with the territory.

Actualized differences, as distinct from abstract differences represented in mathematics and information theory, are characterised by there being something it is like, intrinsically, to undergo those differences, that is, to undergo antagonistic states of opposing forces. All actualized differences undergo this something it is like-ness, but not all are conscious.

It is proposed that a particular kind of activity occurs in human brains that causes our conscious experience. It is a certain dynamic organization of energetic processes having a high degree of differentiation and integration. This organization is recursively self-referential and results in a pattern of energetic activity that blossoms to a degree of complexity sufficient for consciousness.

If consciousness is a physical process, and physical processes are driven by actualized differences of motion and tension, then there is something it is like to undergo actualized differences organized in a certain way in the brain, and this is what we experience — intrinsically.[21]

**Acknowledgements**

I am grateful to the following for discussions and suggestions: Alistair Burleigh, Alan Dix, Chris Doran, Robert K. Logan, Heddwyn Loudon, Chen Song, Galen Strawson, Emmett Thompson, Sander Van der Cruys, and to the reviewers for their comments and criticism. I also wish to acknowledge the support of the Vice Chancellor's Board of Cardiff Metropolitan University.

---

[21] The explanatory principle outlined here might be construed as a form of panpsychism or panexperientialism. My claim is not that consciousness is a fundamental property of nature, universally distributed. Rather, I claim it is a property of all physical systems that there is something it is like, intrinsically, to undergo actualized differences, a certain organization of which causes consciousness.